\documentclass[10pt,a4]{article} 
\usepackage{makeidx}

\newif\ifpdf
\ifx\pdfoutput\undefined
   \pdffalse
\else
   \pdfoutput=1
   \pdftrue
\fi

\ifpdf
  \usepackage[
  pdftex,
  hyperindex,
  pagebackref,
  colorlinks,
  linkcolor=blue,
  menucolor=blue,
  pagecolor=blue,
  urlcolor=blue,
  citecolor=blue,
  pdftitle={The GNSS-R Eddy Experiment II: L-band and optical speculometry for directional sea-roughness retrieval from low altitude aircraft},
  pdfauthor={Olivier Germain},
  pdfcreator={LaTeX}
  ]{hyperref}

  \usepackage{thumbpdf}
  \usepackage{url}
  \usepackage[pdftex]{graphicx}
  \DeclareGraphicsExtensions{.pdf,.jpg,.mps,.png} 

\else

  \usepackage[dvips]{graphicx}
  \DeclareGraphicsExtensions{.eps,.jpg,.mps,.png,.ps}

\fi                          

\usepackage{fancyhdr}
\usepackage{xfig}
\usepackage{multicol}     
\graphicspath{{figures/}}
\usepackage[english]{babel}
\usepackage[]{times}

\catcode`\@=11

\catcode`\@=11
\def\section{\@startsection{section}{1}{\z@}{3.5ex plus 1ex minus
   .2ex}{2.3ex plus .2ex}{\large\bf}}

\def\eqnarray{\let\@currentlabel=\theequation\refstepcounter{equation}
    \global\@eqnswtrue
    \global\@eqcnt\z@\tabskip\@centering\let\\=\@eqncr
    $$\halign to \displaywidth\bgroup\@eqnsel\hskip\@centering
      $\displaystyle\tabskip\z@{##}$&\global\@eqcnt\@ne
       \hfil${{}##{}}$\hfil
      &\global\@eqcnt\tw@ $\displaystyle\tabskip\z@{##}$\hfil
       \tabskip\@centering&\llap{##}\tabskip\z@\cr}
\def\lefteqn#1{\hbox to 4\arraycolsep{$\displaystyle #1$\hss}}

\def\thesection{\arabic{section}.}

\def\appendix{\setcounter{section}{0}
        \def\thesection{\Alph{section}.}
        \def\theequation{\Alph{section}.\arabic{equation}}}

\long\def\@makefntext#1{\parindent 0cm\noindent
\hbox to 1em{\hss$^{\@thefnmark}$}#1}

\newcommand\unmarkfootnote[1]{%
  \begingroup
    \let\@makefntext\noindent
    \footnotetext{#1}%
  \endgroup}

\newcommand{\captionfonts}{\footnotesize}

\makeatletter  
\long\def\@makecaption#1#2{%
  \vskip\abovecaptionskip
  \sbox\@tempboxa{{\captionfonts #1: #2}}%
  \ifdim \wd\@tempboxa >\hsize
    {\captionfonts #1: #2\par}
  \else
    \hbox to\hsize{\hfil\box\@tempboxa\hfil}%
  \fi
  \vskip\belowcaptionskip}
\makeatother   

\begin{document}

\thispagestyle{empty}

\noindent
{\Large \bf The GNSS-R Eddy Experiment II: L-band and Optical Speculometry for Directional Sea-Roughness Retrieval from Low Altitude Aircraft} 

\vspace*{0.3cm}
\noindent 
O. Germain, G. Ruffini, F. Soulat, M. Caparrini  \\
 {\it Starlab, C. de l'Observatori Fabra s/n, 08035 Barcelona, Spain, http://starlab.es} 

\medskip \noindent
B. Chapron \\
 {\it Ifremer,  Technopole de Brest-Iroise BP 70, 29280 Plouzan\'e, France,  http://ifremer.fr}

\medskip \noindent
P. Silvestrin \\
 {\it ESA/ESTEC, Keplerlaan 1, 2200 Noordwijk, The Netherlands,  http://esa.int} 
\medskip


\vspace*{0.3cm}

\section*{Abstract}

We report on the retrieval of directional sea-roughness (the full directional
mean square slope, including MSS, direction and isotropy) through inversion of Global Navigation Satellite System Reflections (GNSS-R) 
and SOlar REflectance Speculometry (SORES)
data  collected during
an experimental flight at 1000 m.
The emphasis is on the utilization of the entire Delay-Doppler Map (for GNSS-R)
or Tilt Azimuth Map (for SORES) in order to infer these directional parameters. 
Obtained estimations are analyzed and compared to Jason-1 measurements
and the ECMWF numerical weather model.    
\medskip\\
{\bf Keywords:} GNSS-R, GPS, Galileo, Speculometry, SORES, sea roughness, Directional Mean Square Slope, Delay-Doppler Map.
\medskip\\

\begin{multicols}{2}

\section{Introduction}

The use of Global Navigation Satellite System (GNSS) signals reflected by the sea surface as a remote sensing tool has generated considerable attention for over a decade~\cite{martin-neira1993,ruffini2001a}.
Among several applications, two classes have rapidly emerged in the community:
sea-surface altimetry, which aims at retrieving the mean sea level like classical radar altimeters do, and sea-surface scatterometry or ``speculometry'' (see below for a justification of this neologism)
 for the determination of sea roughness and near surface wind. This paper addresses the latter application. 


Inferring sea roughness from GNSS-R data requires (i) a parametric description of the sea surface, (ii) an electromagnetic model for sea-surface scattering at L-band and (iii) the choice of a GNSS-R data product to be inverted. 
In the literature, there is quite an agreement on the two first aspects.
It has been recognized that the scattering of GNSS signals can be  modeled as a Geometric Optics process (GO), where the fundamental physical process is the scattering from mirror-like surface elements. This is the reason why we use the term ``speculometry'', from the latin word for mirror, {\it speculo}. Therefore, the most important feature of the sea surface is the statistics of facet slopes at about the same scale as the electromagnetic wavelength ($\lambda$). This is described by 
the bi-dimensional slope probability density function (PDF). 
Under a Gaussian assumption,  three parameters suffice to fully define the sea-surface PDF: the directional mean square slope DMSS$_\lambda$, which results from  the integration of the ocean energy spectrum at wavelengths larger than $\lambda$.
The symbol DMSS$_\lambda$ englobes the three parameters defining the ellipsoidal shape of the slope PDF: scale (total MSS), direction (Slope PDF azimuth) and isotropy (Slope PDF isotropy).  
The GNSS-R scattering model proposed by Zavorotny and Voronovich in~\cite{zavorotny2000} is based on GO, and is, to date, the reference model for the GNSS-R community. 
While for the purposes of specular scattering the sea-surface roughness is parametrized by the directional mean square slope in a direct manner, DMSS$_\lambda$  is rarely emphasized as the geophysical parameter of interest. Instead, most authors prefer to link sea roughness to the near surface wind vector, which is thought to be more useful for oceanographic and meteorological users, but misleading. Indeed, this link requires an additional modeling layer and is an extra source of error. For instance, a wind-driven sea spectra is not suitable for inferring sea surface DMSS$_\lambda$ when swell is present or the sea not fully developed. The connection between DMSS$_\lambda$ and wind is thus modulated by other factors (e.g., swell, fetch and degree of maturity).


Usually, for technical reasons, the product inverted in GNSS-R speculometry is a simple Delay Waveform, that is, a 1D delay map of the reflected signal amplitude. Using a single GNSS emitter, the wind speed can be inferred assuming an isotropic slope PDF (i.e., the PDF's shape is a circle)~\cite{garrison2002,cardellach2003,komjathy2000}.
Attempts have also been made to measure the wind direction by fixing the PDF isotropy to some theoretical value (around 0.7) and using at least two satellites reflections with different azimuths~\cite{zuffada2000a,garrison2003}.
As investigated in the frame of the ESA OPPSCAT project (see~\cite{oppscat2000} and \cite{ruffini2000a}), it is nonetheless possible to work on a product of higher information content: the Delay-Doppler Map (DDM),  a 2D delay-Doppler map of the reflected signal amplitude.
The provision of an extra dimension opens the possibility to perform the full estimation of the DMSS$_\lambda$. 
In~\cite{elfouhaily2002a}, Elfouhaily {\it et al.} developed a rapid but sub-optimal method based on the moments of the DDM to estimate the full DMSS$_\lambda$: this approach neglects the impact of the bistatic Woodward Ambiguity Function modulation of the Delay Doppler return.


The present paper was motivated by a recent experiment conducted by Starlab for the demonstration of GNSS-R altimetry. The altimetric aspects are reported elsewhere~\cite{ruffini2003}. 
We note that the configuration of the flight was not optimized for speculometry:
from 1000~m altitude, the sea-surface reflective area is essentially limited by the PRN C/A code, and the glistening zone is coarsely delay-Doppler mapped.
In addition to the GNSS-R experiment, high resolution optical photographs of sun glitter were also taken, providing the SORES dataset (SOlar REflectance Speculometer).
Since the classic study of Cox and Munk~\cite{cox1954}, it is well known that sea-surface DMSS$_{Opt}$ can be inferred from such data. The  availability of optical photographs thus provided us with an extra source of colocated information.
Because there is a strong similarity between products ---DDM for GNSS-R and the Tilt Azimuth Map (TAM) for SORES--- and models ---GO--- the same inversion methodology can be applied to both datasets.

The goal of the paper is to investigate the full exploitation of the bidimensional GNSS-R DDM  and  SORES TAM 
products to infer the set of three DMSS$_\lambda$ parameters.  The driver of the study has been the exhaustive exploitation of the information contained in those 2D products. Consequently, the proposed approach relies on a least-square fit of speculometric models to datasets. 
We first describe in details the collected datasets and the associated pre-processing. Then, we present the speculometric models needed to infer data together with the inversion scheme. Finally, we provide the estimation results and discuss their coherence with other sources of data.

\section{Dataset collection and pre-processing}

\begin{figure*}[t!]
\centering
\includegraphics[width=8cm]{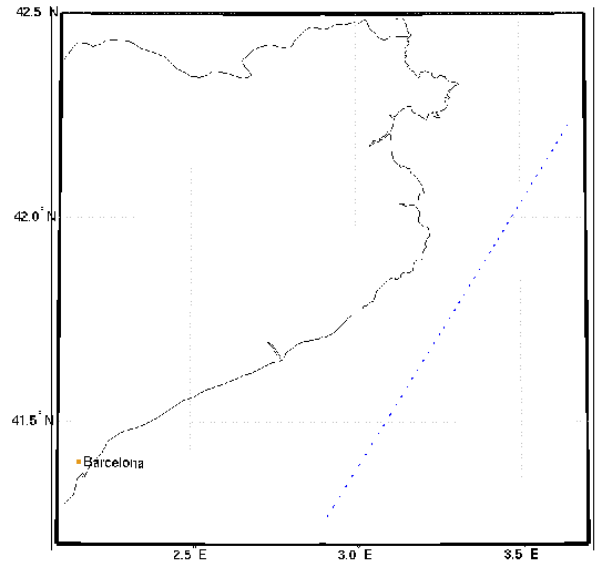}
\caption{
Map of the aircraft track divided into 46 10-second arcs.
}
\label{the maps}
\end{figure*}

\begin{figure*}[t!]
\centering
\begin{tabular}{cc}
\includegraphics[width=6cm]{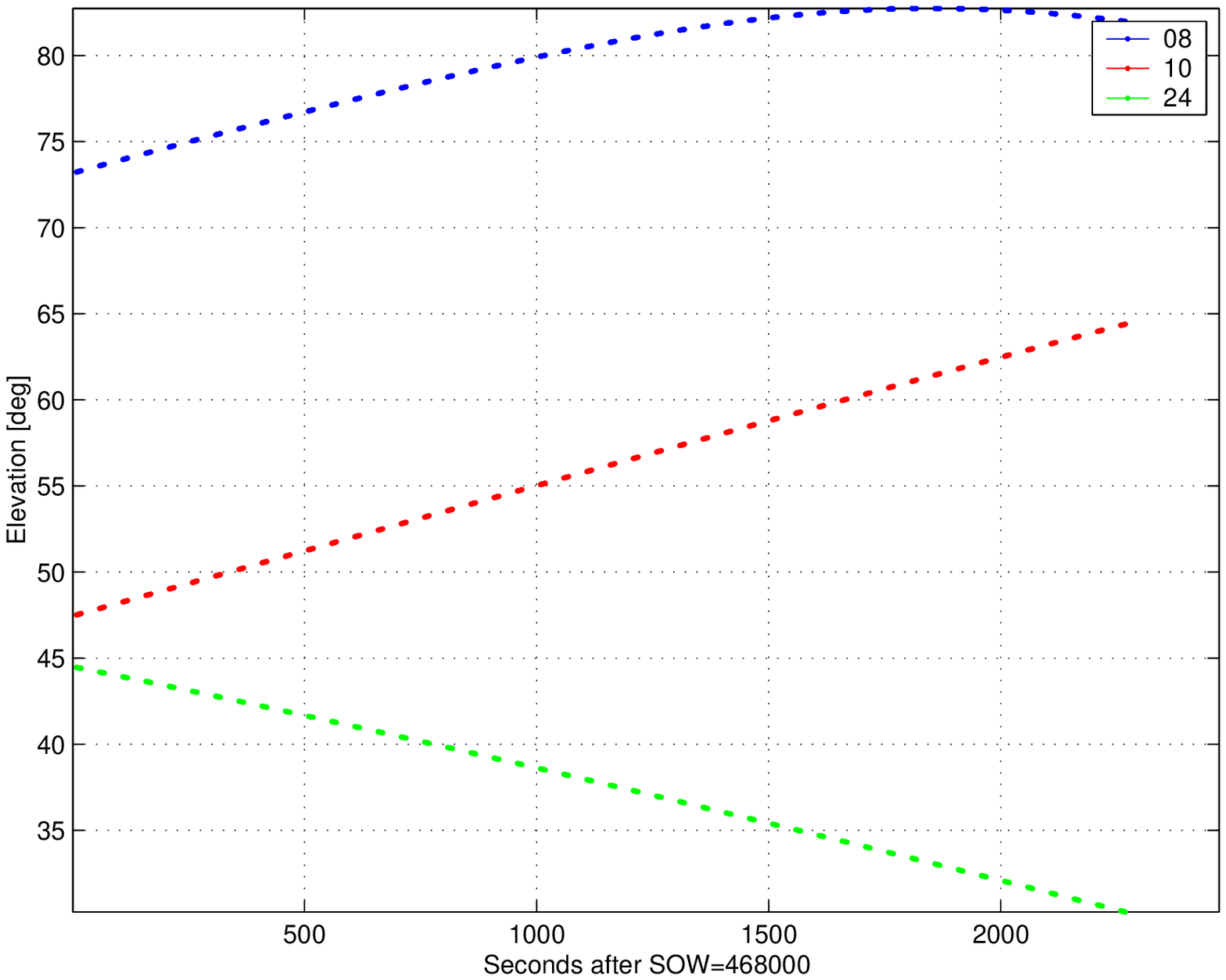} &
\includegraphics[width=6cm]{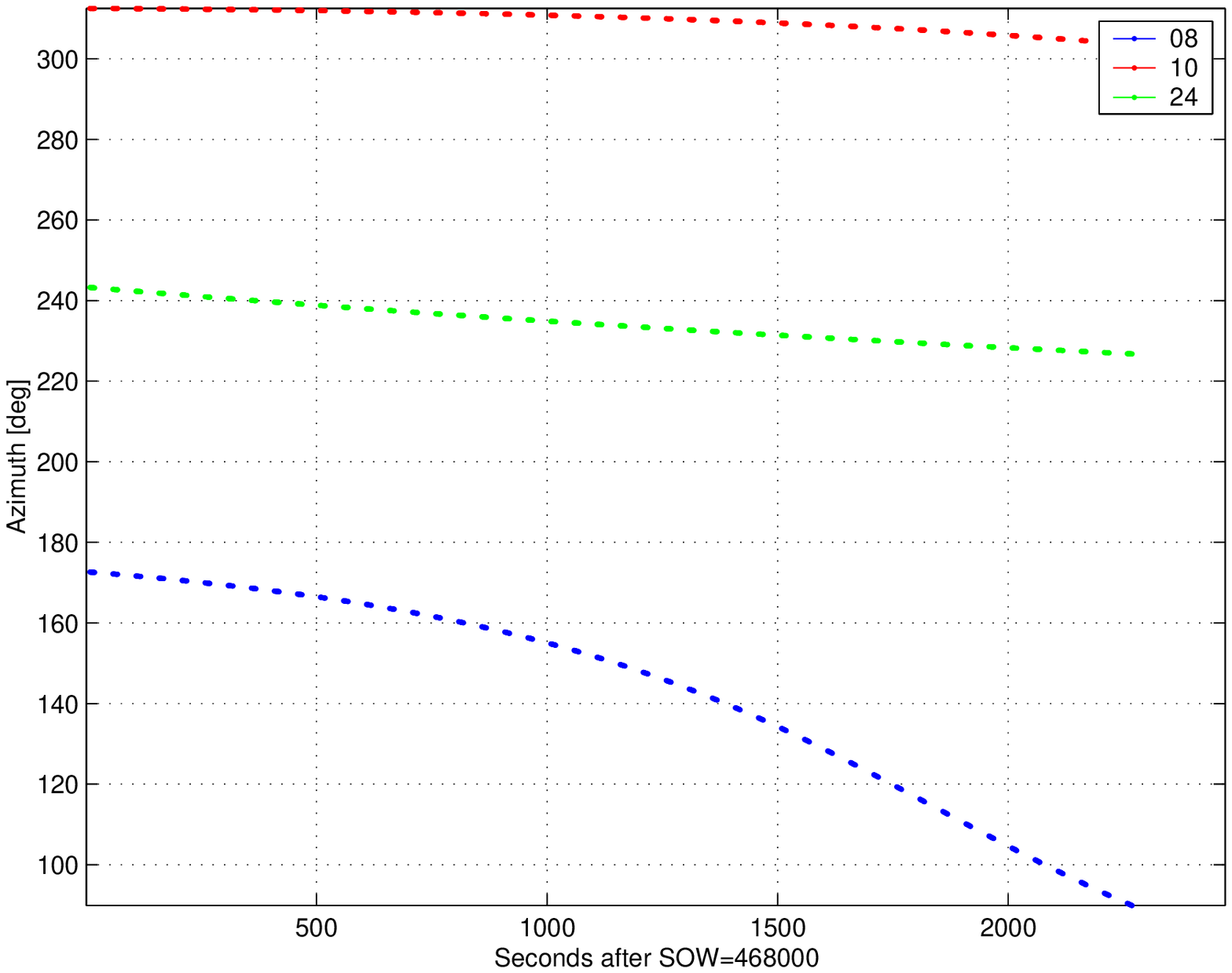}
\end{tabular}
\caption{Elevation and Azimuth of three GPS satellites in view during the 46 10-second arcs.}
\label{sat_in_view}
\end{figure*}

The campaign took place Friday September 27th 2002 around 10:00 UTC, along the Catalan coast from Barcelona (Spain) up to the French border.
An aircraft overflew the Jason-1 track~187 at 1000~m along 150~km and gathered 1.5 hours of GPS-R raw signals (see~\cite{soulat2003d} for more details).
Since it would have been computationaly too expensive to process the full track, it was divided into 46 10-second arcs (each spanning roughly 500 meters), sampled every 50 s (see Figure~\ref{the maps}). The first arc started at GPS Second Of the Week 468008.63. 
The aircraft
kinematic parameters were kept close to the nominal values specified in the mission plan: altitude=1000~m, speed=50~m/s and heading from North=30$^o$.
We have selected three GPS satellites in optimal view during the experiment whose elevation and azimuth are given in Figure~\ref{sat_in_view}.


The raw GNSS-R data were recorded using the GPS reflection equipment
provided by ESA. Specifically, the GPS direct and reflected signals were 1-bit sampled and stored a rate of 20.456 Mbit/s.
The pre-processing step consisted in performing a delay-Doppler Pseudo Random Noise (PRN) code despreading to coherently detect the direct signal (from GPS emitter) and the reflected signal (scattered by sea-surface).   
We used the Starlab in-house software to produce three DDMs time-series (one per PRN), sampled into 46 arcs of 10 seconds each.
The general processing strategy was to track the delay-Doppler of direct signal and then compute DDMs for both direct and reflected signals. These DDMs actually represent the filtered electromagnetic field of incoming signals, as processed with delay-Doppler value slightly different from those corresponding to the specular point.
The coherent integration time was set to 20~ms to ensure a Doppler resolution of 50~Hz.
The delay map spanned 80 correlation lags (i.e. +/- 1.95 $\mu$s) with a lag step of 48.9 ns, while the Doppler range spans -200~Hz to 200~Hz with a step of 20~Hz.
Incoherent averaging was applied to each arc (the accumulation time was set to 10~s). 
This process aimed at reducing both thermal and speckle noise by a factor of $\sqrt{500}$.
At the end, the GNSS-R product for one PRN and one arc was an average amplitude delay-Doppler map of size 81$\times$21.     


The SORES photographs were taken from time to time along the track when the roll, pitch and yaw angles of the plane were negligible. The camera was a Leica dedicated to aerial photography. An inertial system (by Applanix) provided the time-tagged position for each snapshot. 
The film was a panchromatic Aviphot Pan 80. The focal length was 15.2 cm, and the photographic plate had an area of 23$\times$23 cm$^2$. The aperture angle was consequently 74.2$^o$. The observed area was a square with area 1.124$\times$1.124 km$^2$. The exposure time was fixed to 1/380 and aperture to f/4.
The silver photographs were scanned and digital images were averaged to 400$\times$400 squared pixels, in order to be easily processed.

\section{Models and inversion}

\subsection{Directional mean square slope}

As discussed above, forward scattered signals at short wavelengths (optical but also L-band) are mostly sensitive to the specular scattering mechanism.
Therefore, the strongest sea-surface signature in received signal is expected to be due to  facet slope statistics at the relevant scales. The 2D slope probability density function (PDF) Gaussian model is given by
\begin{equation}
{\cal P}(s) =
\frac{
e^{
-\frac{1}{2}
s  
^{\dag}
K^{-1}
s
}
}{2 \pi \sqrt{{\mbox{det}}(K)}},
\end{equation}
where $s^{\dag}= \left[ \partial \zeta / \partial x ~~ \partial \zeta / \partial y\right]$ stands for the vector of directional slopes in some frame
$xy$ and $K$ is the matrix of slope second order moments.
The $xy$ frame mapped on sea-surface is defined as follows: it is centred on the specular point and has its $x$ axis aligned with the
Transmitter-Receiver line.
Mean-square slopes along major and minor principal axes are often referred to as MSS up-wind ($mss_u$)
and MSS cross-wind ($mss_c$) respectively. The $K$ matrix is then obtained via a simple rotation,
\begin{equation}
K =
R_{\psi}
\cdot
\left[
\begin{array}{cc} mss_u & 0 \\ 0 & mss_c \end{array}
\right]
\cdot
R_{-\psi},
\end{equation}
where $R_\psi$ is the usual rotation matrix of angle $\psi$, the angle between the $x$-axis and the slope principal axis.          

\begin{figure*}[t!]
\centering
\includegraphics[width=8cm]{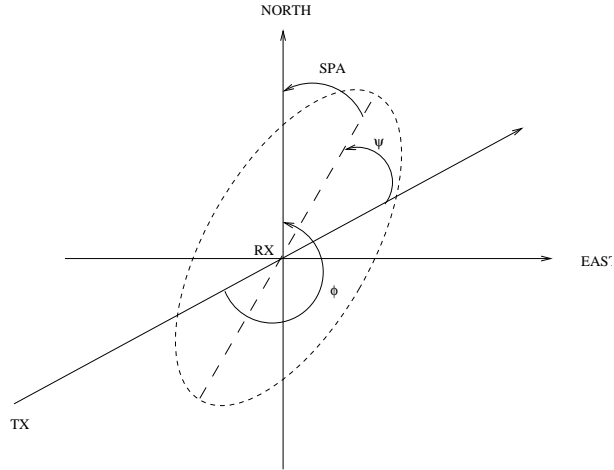}
\caption{Sketch of the slope PDF and related frames.
}
\label{geometry}
\end{figure*}

Thus, $mss_u$, $mss_c$ and $\psi$ are the three geophysical parameters we wish to estimate.
They can be thought of as the three parameters of an ellipse (see Figure~\ref{geometry}) representing
the slope PDF mapped on sea-surface.
In the following, we will consider the equivalent set of parameters:
\begin{itemize}
\item Total MSS, defined as: $2 \sqrt{mss_u . mss_c}$. This magnitude is actually proportional to ellipse area and
can be interpreted in terms of wind speed, based on
a particular wind-driven sea-surface spectrum like the Elfouhaily's spectrum~\cite{elfouhaily1997}.
\item Slope PDF azimuth (SPA), defined as the direction of semi-major axis with respect toNorth. As shown by Figure~\ref{geometry}, this angle is $ \pi + \Phi - \psi$, if $\Phi$ i
s the satellite azimuth from North.  
\item Slope PDF isotropy (SPI), defined as $mss_c / mss_u$. When SPI=1, the slope PDF is isotropic and the glistening zone is circular. Low values
of SPI indicate a highly directive PDF. Typically, SPI is expected to be around 0.65 for well
developed sea-surface.
\end{itemize}

\subsection{GNSS-R speculometric model}

The classical GNSS-R bistatic
radar equation~\cite{zavorotny2000} links the average GNSS-R power return
to sea-surface slope PDF.
Processing the raw signal with various
delay-Doppler values $(\tau,f)$, a DDM is computed whose theoretical expression is:
\begin{eqnarray}
P(\tau,f) & = &
\int  dx dy \,
\frac{G_r}{R_t^2 R_r^2} \cdot
\sigma ^0 \cdot \nonumber \\
&&
\chi^2 \left( \tau_m-\tau_c-\tau,f_m-f_c-f \right) 
\label{radar_eq}
\end{eqnarray}
where
$G_r$ is the receiver antenna pattern,
$R_t$ and $R_r$ are the distances from generic point on sea-surface to transmitter
and receiver,
$\sigma^0$ is the reflectivity, 
$\chi$ is the Woodward Ambiguity Function (WAF, see~\cite{ruffini2001a}),
$\tau_m(x,y)$ and $f_m(x,y)$ are the delay-Doppler mapping on sea-surface
and $\tau_c$ and $f_c$ are delay-Doppler centers.
To first order, the reflectivity is proportional to the slope PDF:
\begin{equation}
\sigma^0  =
\pi |{\cal{R}} |^2 \frac{q^4}{q_z^4}
~{\cal P} \left( \frac{-q_x}{q_z},\frac{-q_y}{q_z} \right) ,
\label{eq_sigma0}
\end{equation}
where $(q_x,q_y,q_z)$ is the scattering vector and $|{\cal{R}} |^2 = 0.65$ is the specular Fresnel coefficient.

The presence of thermal noise biases the value of average power return. 
Hence, the average amplitude of the DDM can by modeled by 
\begin{equation}
A(\tau,f) = \sqrt{\alpha . P(\tau,f) + b},
\end{equation}
where $b$ stands for the bias in power.
In particular, this effect is visible in the early-delay domain of the DDM:
for delays lower than one-chip, the DDM amplitude has a no null
value,
often called "grass level".
As we do not have a calibrated model an overall scaling
parameter $\alpha$ is also needed in the model.


To sum up, the model features three parameters of interest and four ``nuisance parameters'':
\begin{itemize}
\item the DMSS$_\lambda$, characterizing the Gaussian slope PDF: total MSS, isotropy (SPI)
and azimuth (SPA),
\item the DDM delay-Doppler centers: $\tau_c$ and $f_c$,
\item overall scaling parameter: $\alpha$,
\item grass level: $b$.
\end{itemize}
Other parameters required to run the forward model are recalled in Table~\ref{theparam}.
\begin{table*} [t!]
\begin{center}
\begin{tabular}{|l|l|}
\hline
\bf{Geometry} &
\begin{tabular}{ll}
Aircraft: & Altitude, speed and heading provided at 1 Hz \\
Satellite: & Elevation and azimuth provided at 1 Hz \\
\end{tabular}
\\ \hline
\bf{Instrument} &
\begin{tabular}{ll}
Antenna Pattern: & measured in anechoic chamber  \\
Band:  & GPS L1 (19 cm)\\
GPS Code: & C/A \\
\end{tabular}
\\ \hline
\bf{Processing} &
\begin{tabular}{ll}
Integration Time: & 20 ms \\
Accumulation Time: & 10 s \\
Doppler span: & [ -200 Hz , 200 Hz ], 20 Hz step \\
Delay span: & [-40 samples, 40 samples], 1 sample step \\
\end{tabular}
\\ \hline  
\end{tabular}
\end{center}
\caption{
Overview of the parameters necessary for running the DDM forward model.
}
\label{theparam}
\end{table*}

\subsection{SORES speculometric model}

\begin{figure*}[t!]
\centering
\includegraphics[width=11cm]{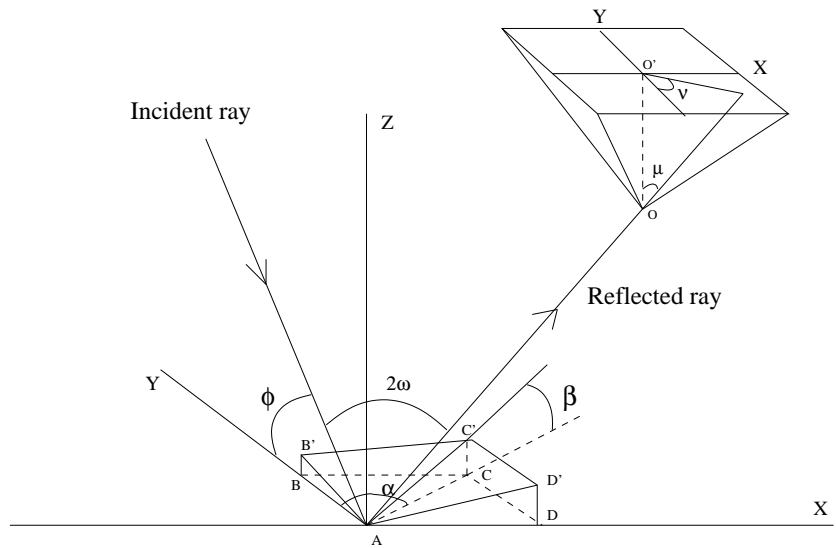}
\caption{Geometry of the SORES experiment.}
\label{sores_geo}
\end{figure*}

To date, results derived from the glitter-pattern of reflected sunlight as photographed by Cox and Munk in 1951 remain the most reliable direct measurements of wind-dependent slope statistics. 
As explained in their well-documented report~\cite{cox1956}, the sea-surface can be gridded with a Tilt ($\beta$) Azimuth ($\alpha$)  Mapping  of the small facet slopes.
These are just a polar parametrization of the ($s_x$,$s_y$) slopes discussed in the previous section:
\begin{equation}
\left\lbrace   
\begin{array}{ccc}
s_x & = & \cos \alpha \cdot \tan \beta , \\  
s_y & = & \sin \alpha \cdot \tan \beta . 
\end{array}
\right.
\end{equation}
The link between the small facet slope PDF ${\cal P}$ and the intensity in the photograh  $I_m$ is given by
\begin{eqnarray} \label{Imodel2}
I_m(\alpha,\beta) &=& A_0 \cdot {\cal P}(\alpha,\beta) \cdot f(\alpha,\beta,\phi) \nonumber \\
& & + K \cdot I_b(\alpha,\beta),
\end{eqnarray}            
where $I_b(\alpha,\beta)$ is the intensity of the picture background (i.e. far from the sun glint), $K$ and $A_0$ are multiplicative constants  and $f$ is a transfer function,
\begin{equation}
f(\alpha,\beta,\phi) = {\rho (1-\rho)^3 \sin \phi \cos^3 \mu \over \cos^3 \beta \cos \omega},
\end{equation} 
with $\phi$: sun elevation, $\rho$: coefficient of reflection and $\mu$, $\omega$: two angles shown on Figure~\ref{sores_geo}.

\begin{figure*}[t!]
\centering
\includegraphics[width=11cm]{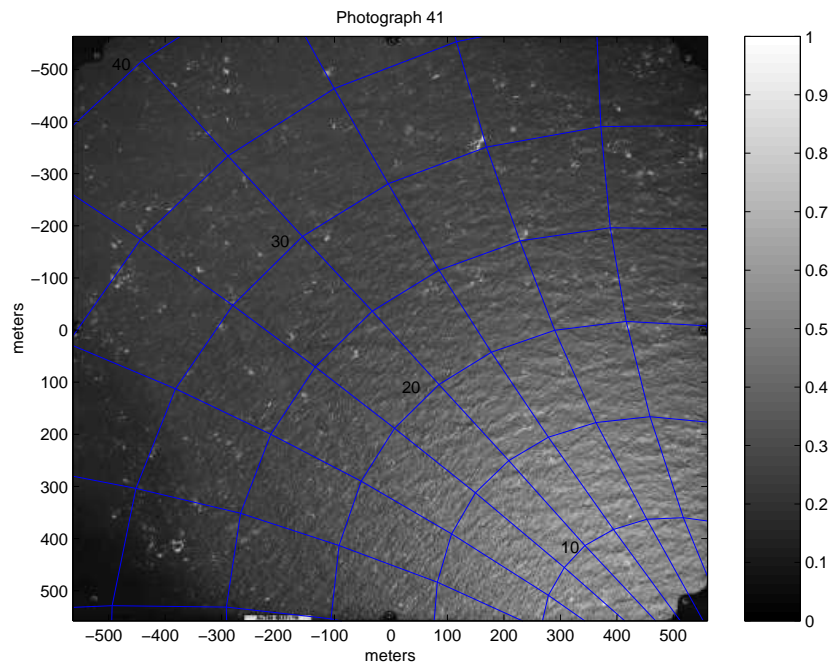}
\caption{Tilt Azimuth Mapping overlaid on a SORES photograph.}
\label{data_photo41}
\end{figure*}  

             
The pixel intensity on the image comes principally from the additive contribution of sunglint and reflected skylight. The sunlight scattered
by particles beneath the sea surface is assumed negligible and is not considered here. A model has been developed to remove reflections of sky radiance from the glint.
The approach consists in considering each sea surface facet specular because, for a given location of the receiver, there always exists a ``source'' in the sky  satisfying the specular reflection condition. Let's consider the cell $(\alpha_i,\beta_i)$ of the TAM. It corresponds to the slope components required to reflect the solar rays onto the camera. The radiance of the sea surface due to reflected skylight in the cell $(\alpha_i,\beta_i)$ can thus be simply modeled by the integration of intensity over all the azimuths $\alpha$ and tilts $\beta$ except the azimuth $\alpha_i$ and the tilt $\beta_i$ of the corresponding cell.

\subsection{Inversion scheme}

\begin{figure*}[t!]
\centering
\begin{tabular}{cc}
\includegraphics[width=8.1cm]{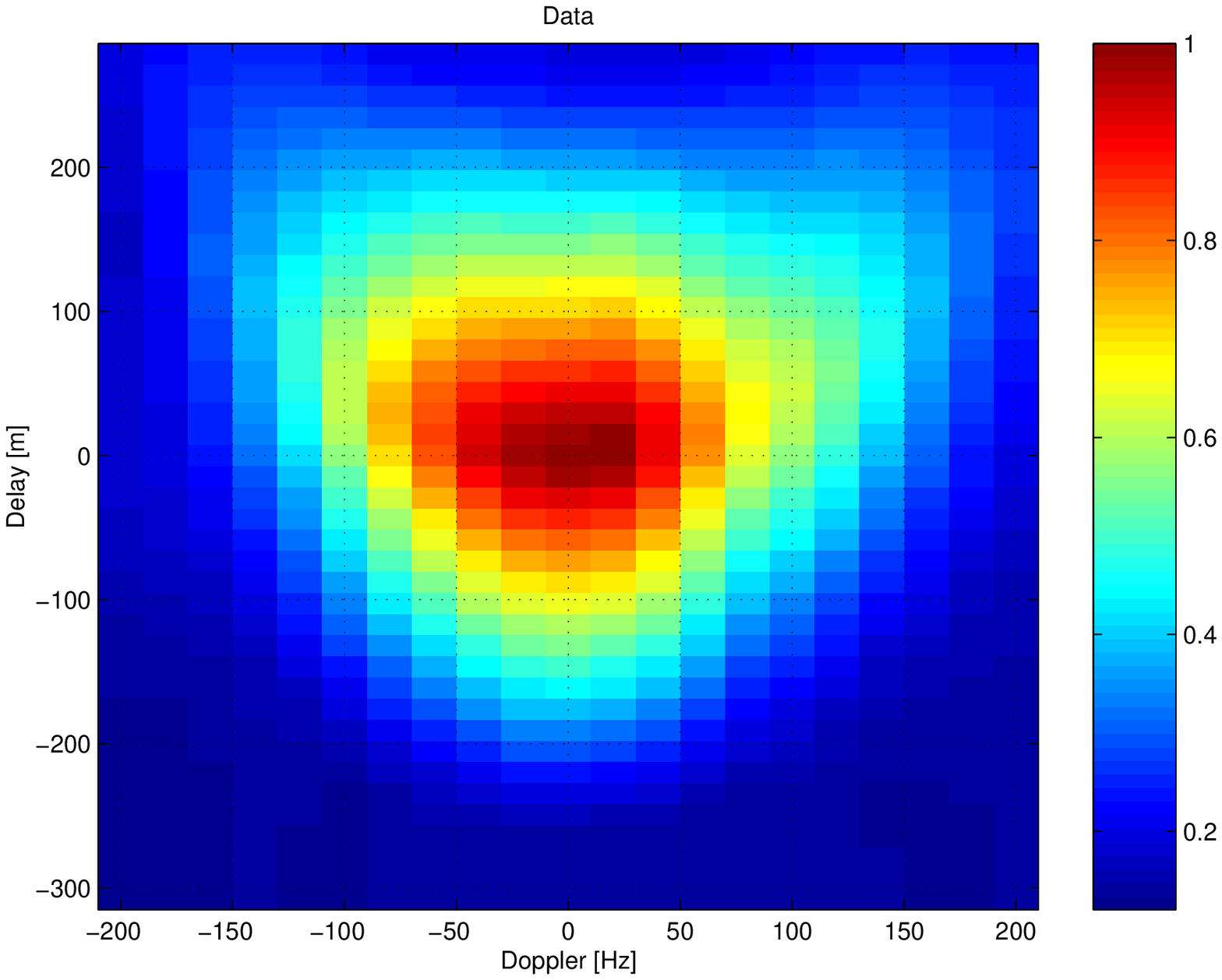} &
\includegraphics[width=8.1cm]{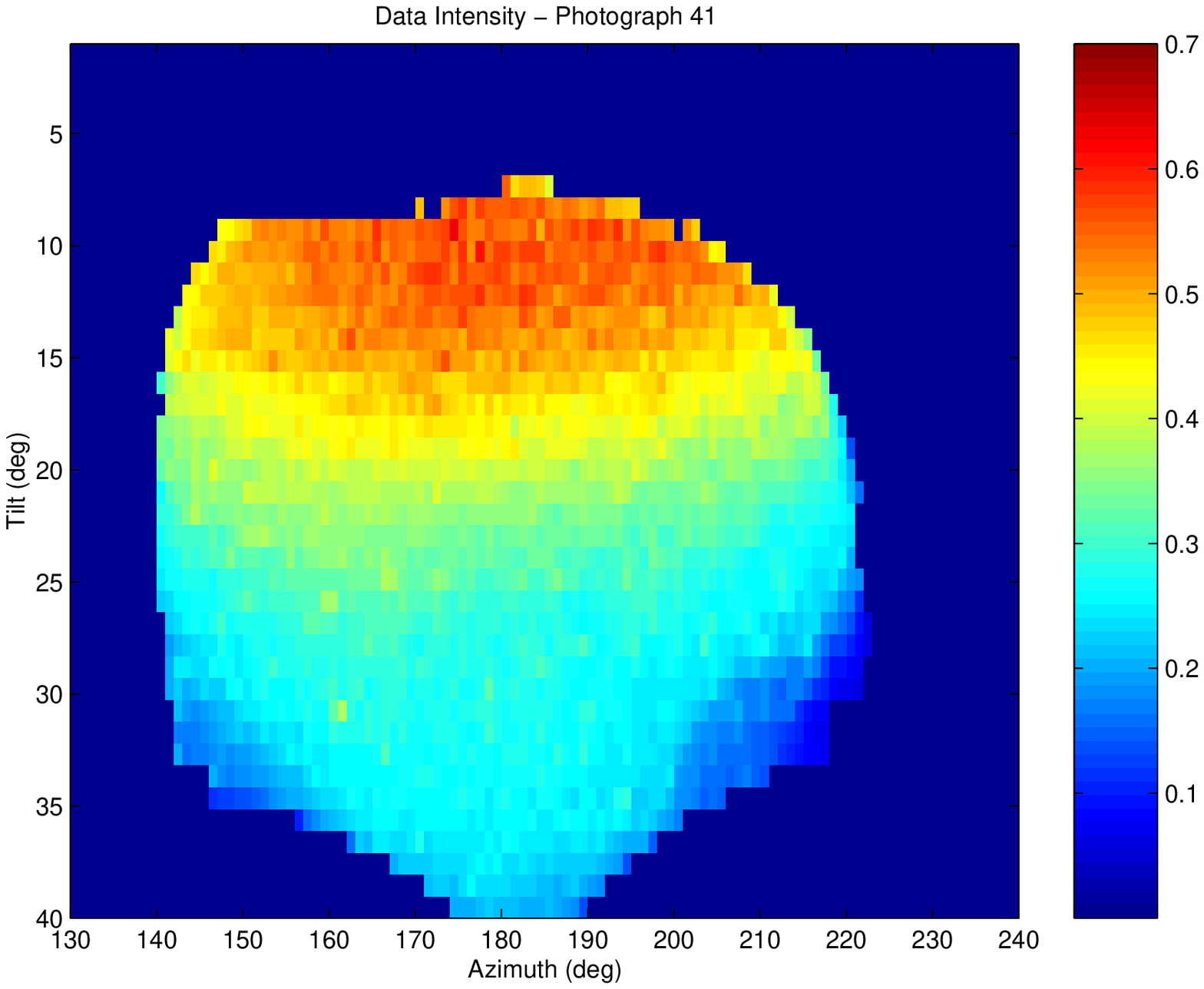} \\
\includegraphics[width=8.1cm]{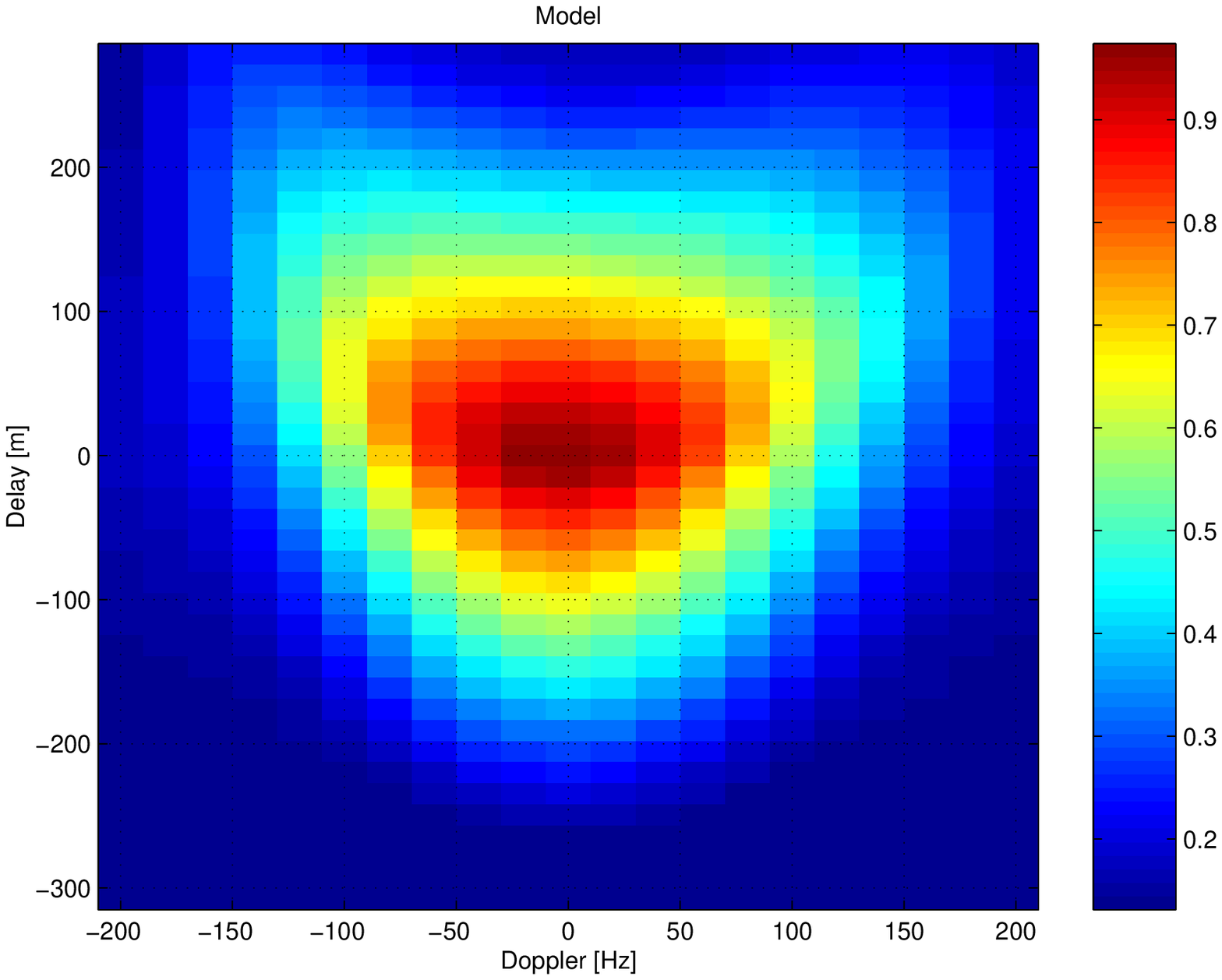} &
\includegraphics[width=8.1cm]{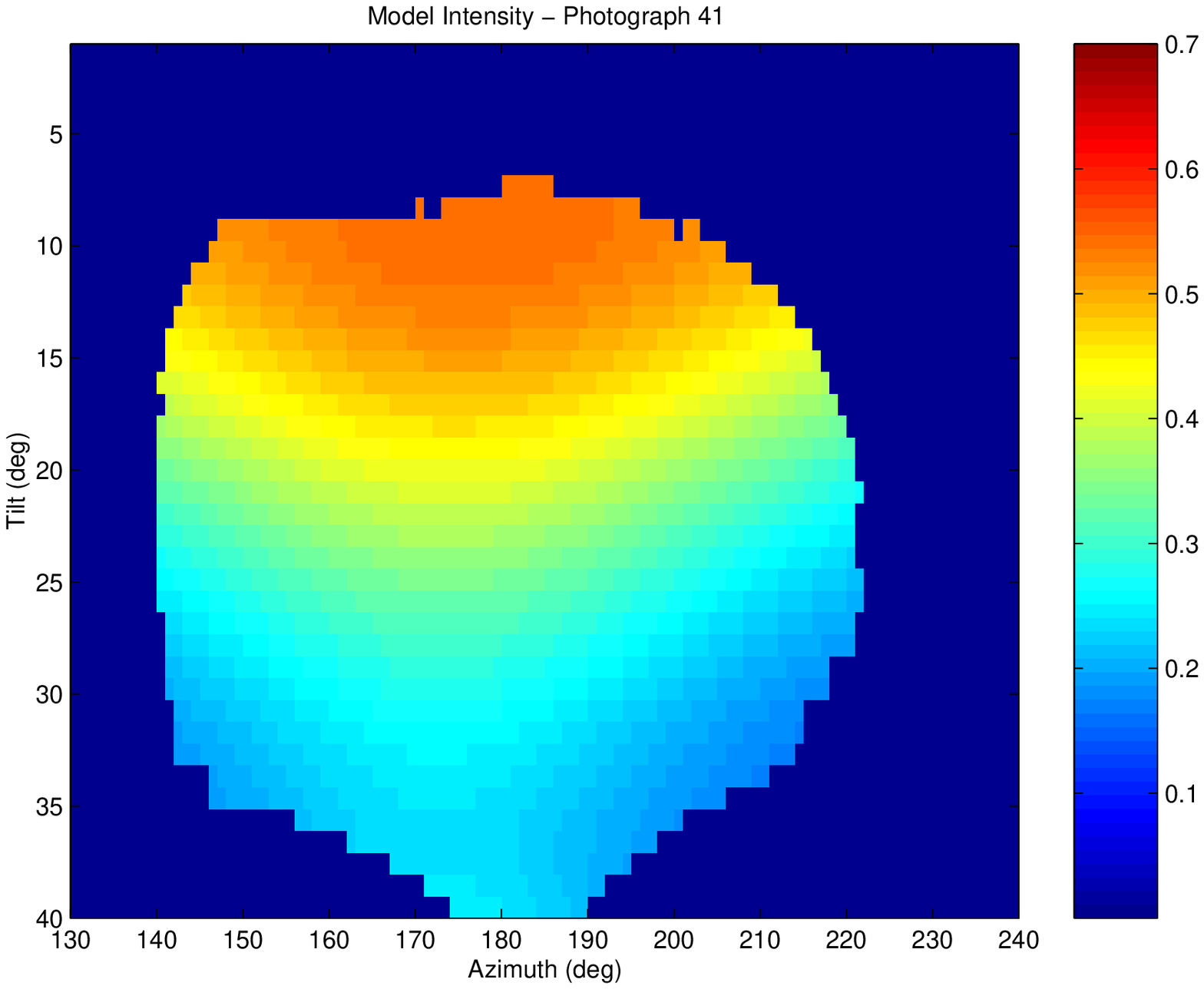} \\
\includegraphics[width=8.1cm]{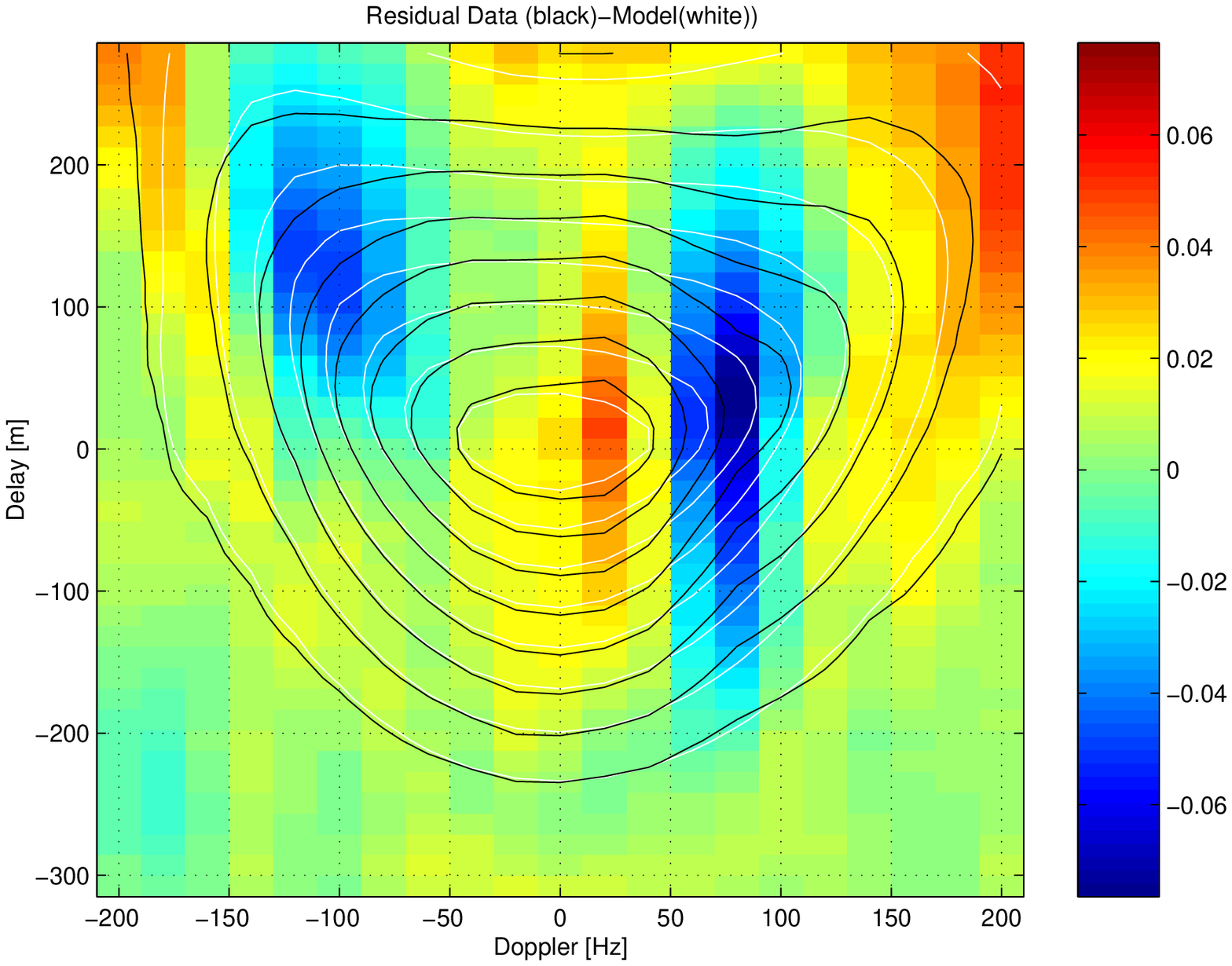} &
\includegraphics[width=8.1cm]{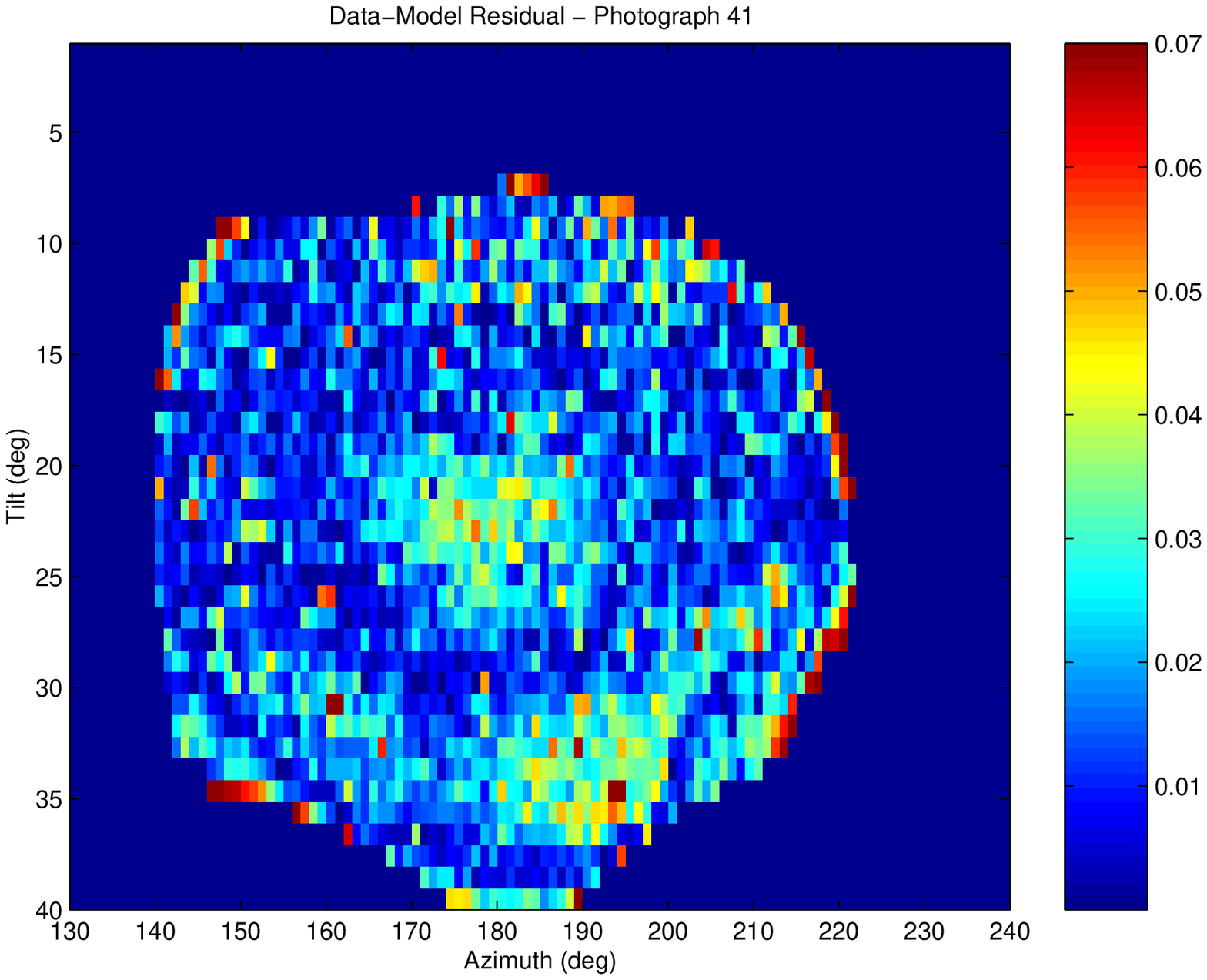}
\end{tabular}
\caption{Examples of data products and their best-fit models.
{\bf First column:} GNSS-R Delay-Doppler Map (PRN08, arc 01).
{\bf Second column:} SORES Tilt-Azimuth Map (photograph 41).
{\bf First row :} Data.
{\bf Second row :} Best-fit Model.
{\bf Third row :} Data-Model Residual.
}
\label{bestfit}
\end{figure*}

Inversion was
performed through a minimization of the root mean square difference between model and data products (i.e., DDMs for GNSS-R and TAMs for SORES). 
Numerical optimization was carried out with a steepest-slope-descent algorithm (Levenberg-Marquardt type adjustment).
The three DMSS$_\lambda$ as well as nuisance parameters
($\tau_c$, $f_c$ and $\alpha$ for the DDMs and $A_0$ and $K$ for the TAMs) were jointly estimated
in an iterative manner:
nuisance parameters (as a first step) and DMSS$_\lambda$ (as a second step) were successively optimized, repeating this two-step sequence until convergence. 
Figure~\ref{bestfit} gives qualitative examples of fit results for DDM and TAM.

\section{Results and discussion}

\begin{figure*}[t!]
\centering
\begin{tabular}{c}
\includegraphics[width=8.7cm]{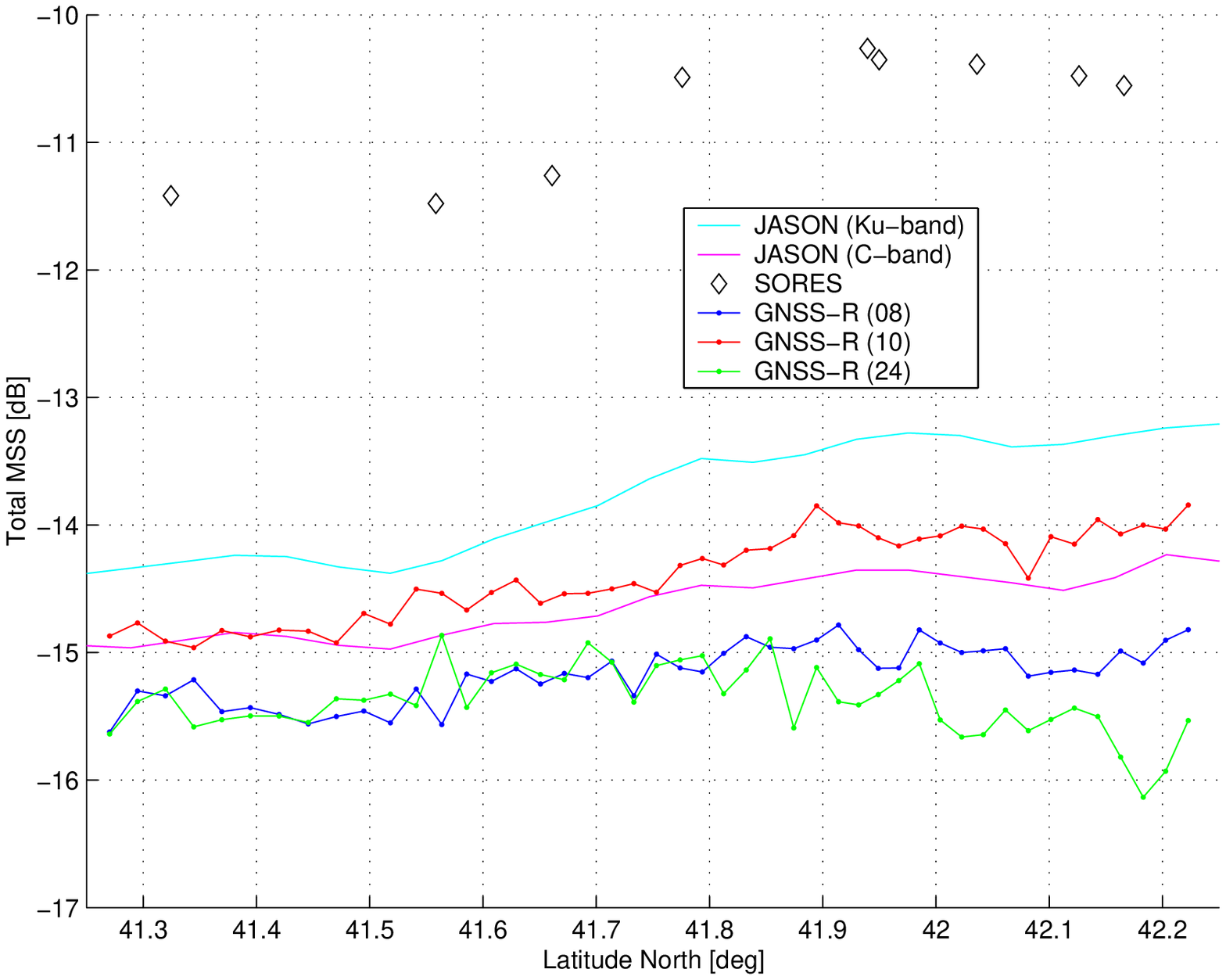} \\
\includegraphics[width=8.7cm]{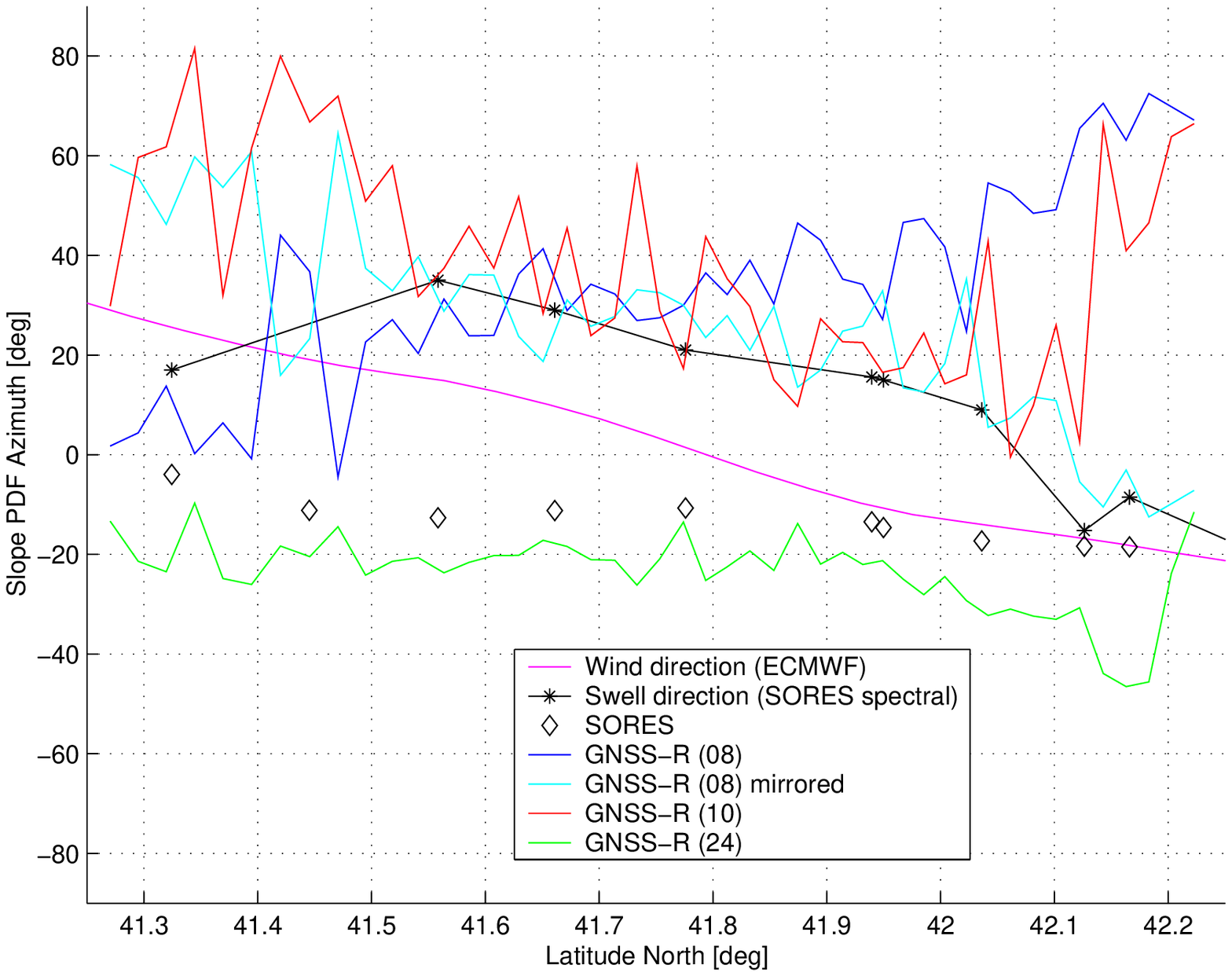} \\
\includegraphics[width=8.7cm]{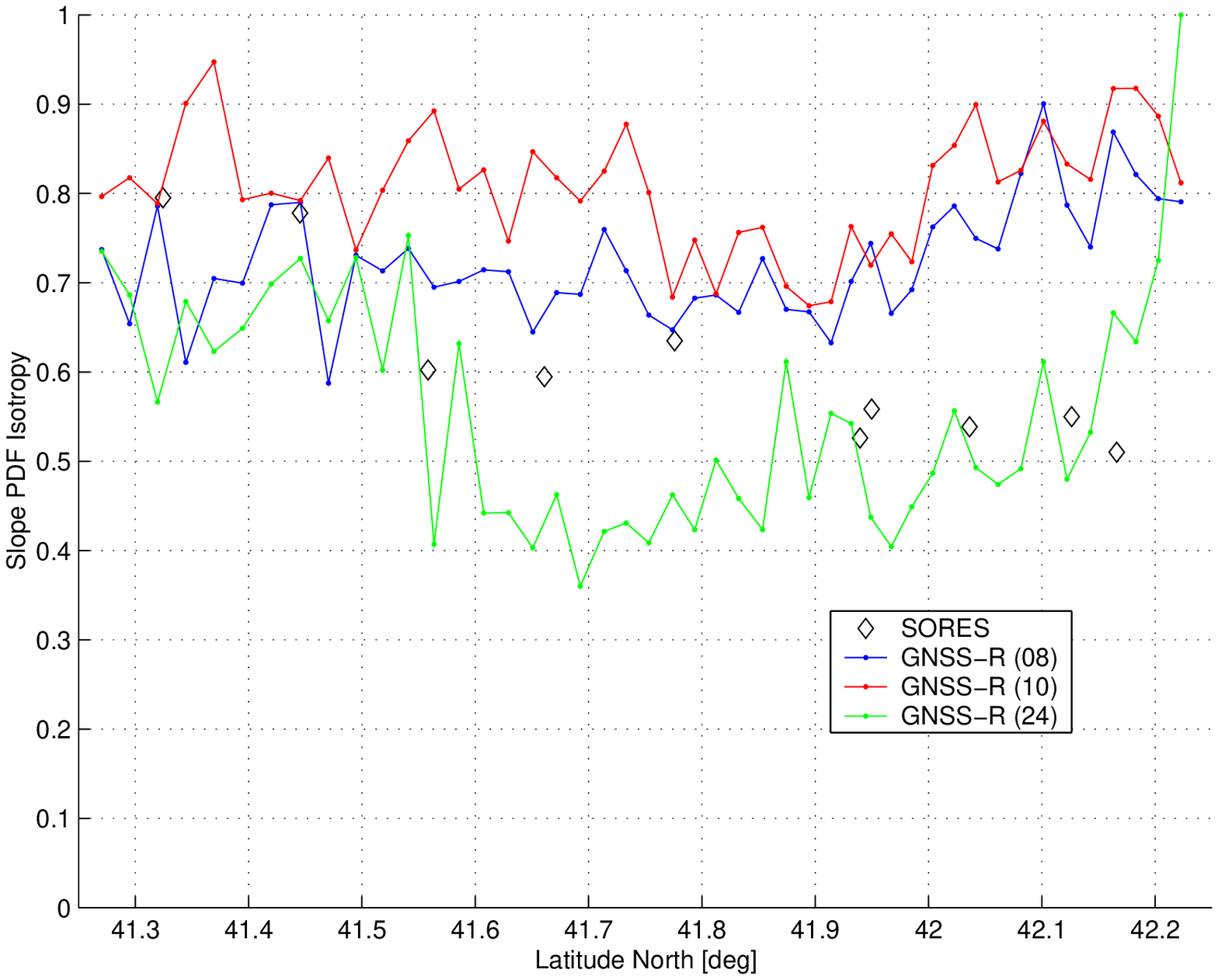}
\end{tabular}
\caption{
DMSS$_\lambda$ estimated along the aircraft track.
{\bf First row: } Total MSS (in dB).
{\bf Second row: } Slope PDF Azimuth.
{\bf Third row: } Slope PDF Isotropy.
}
\label{COMPAR}
\end{figure*}

Figure~\ref{COMPAR} shows total MSS, Slope PDF Azimuth and Slope PDF Isotropy along the aircraft track
between latitudes 41.2$^o$ and 42.2$^o$, as estimated by SORES and GNSS-R.
Other sources of information are also shown for comparison:
\begin{itemize}
\item total MSS for C- and Ku- bands, derived from the Jason-1 $\sigma^o$ measurements at 1 Hz sampling (7 km) and 20 km resolution,
\item wind direction provided by the ECMWF numerical weather model, with accuracy of 20$^o$, and
\item swell direction derived from a spectral analysis of SORES images (the periodic pattern of long waves is indeed clearly observed on the photographs).
\end{itemize}

\subsection{Total MSS}

The total MSS has been plotted in log-scale in order to compare different frequency measurements more easily.
The common trend for all bands is the increase of slope variance 
with latitude until  a relative plateau is reached.
Measurements of PRN08 and 24 show good agreement while PRN10 seems to be somewhat up-shifted.
As expected, we observe that the level and dynamic of MSS decrease with longer wavelength:
Optical, Ku, C and L band, in this order.
Nevertheless, 
the level and dynamic range of GNSS-R plots (especially PRN10) seem a bit large for L-band measurements, when compared to C-band.
Note however that Jason-1's MSS have been obtained through the relationship
$MSS = \kappa / \sigma^o$,  $\kappa$ being an empirical parameter accounting for
calibration offsets. 
Unfortunately, the uncertainty on $\kappa$ makes the absolute levels of Jason-1's plots very doubtful. 
Here, as an illustration purpose, we have set $\kappa$=0.45 and $\kappa$=0.95 for C- and Ku-band respectively.

\subsection{Slope PDF Azimuth}

Using a single DDM, the estimation of SPA is degenerate in two particular cases: when the transmitter is 
at zenith or when the
receiver moves torwards the transmitter~\cite{cardellach2000}. In these two cases, the Delay-Doppler lines
that map the glistening zone are symmetric
around the receiver direction. Hence, one cannot distinguish between a slope PDF and its mirror
image about the receiver direction.
Here, PRN08 has its elevation comprised between 74 and 83 degrees. It is then very likely that the
SPA estimated for this PRN is degenerate. For this reason, we have added on the plot the mirror image of the SPA about the receiver direction (30$^o$).
We also note that the azimuth of PRN10 decreases down to 230$^o$
at the end of the track, quite close from  210$^o$, the complementary of the receiver's direction. 


According to ECMWF data and SORES spectral analysis, wind and swell were slightly misaligned.
PRN08 (or its mirror image) matches very well the swell direction and so does PRN10 along
most of the track. This result underlines the fact that GNSS-R is not sensitive
to wind only and that swell has a strong impact too.
PRN24 has a different behaviour, in line with SORES. These two measurements  agree relatively
better with
wind direction, although a discrepancy of 30$^o$ is observed at the beginning of the track.

\subsection{Slope PDF Isotropy}

It is worth remembering that Elfouhaily's wind-driven spectrum predicts a SPI value of 0.65, hardly sensitive to wind speed.  
Here we note that SPI varies quite significantly along the track for both GNSS-R and SORES.
The important departure observed from the 0.65 nominal value is probably a signature
of an under-developed sea and the presence of strong swell.
Further research should be undertaken in order to better understand the potential information contained
in this product.

\subsection{Link to wind speed}

\begin{figure*}[t!]
\centering
\includegraphics[width=15cm]{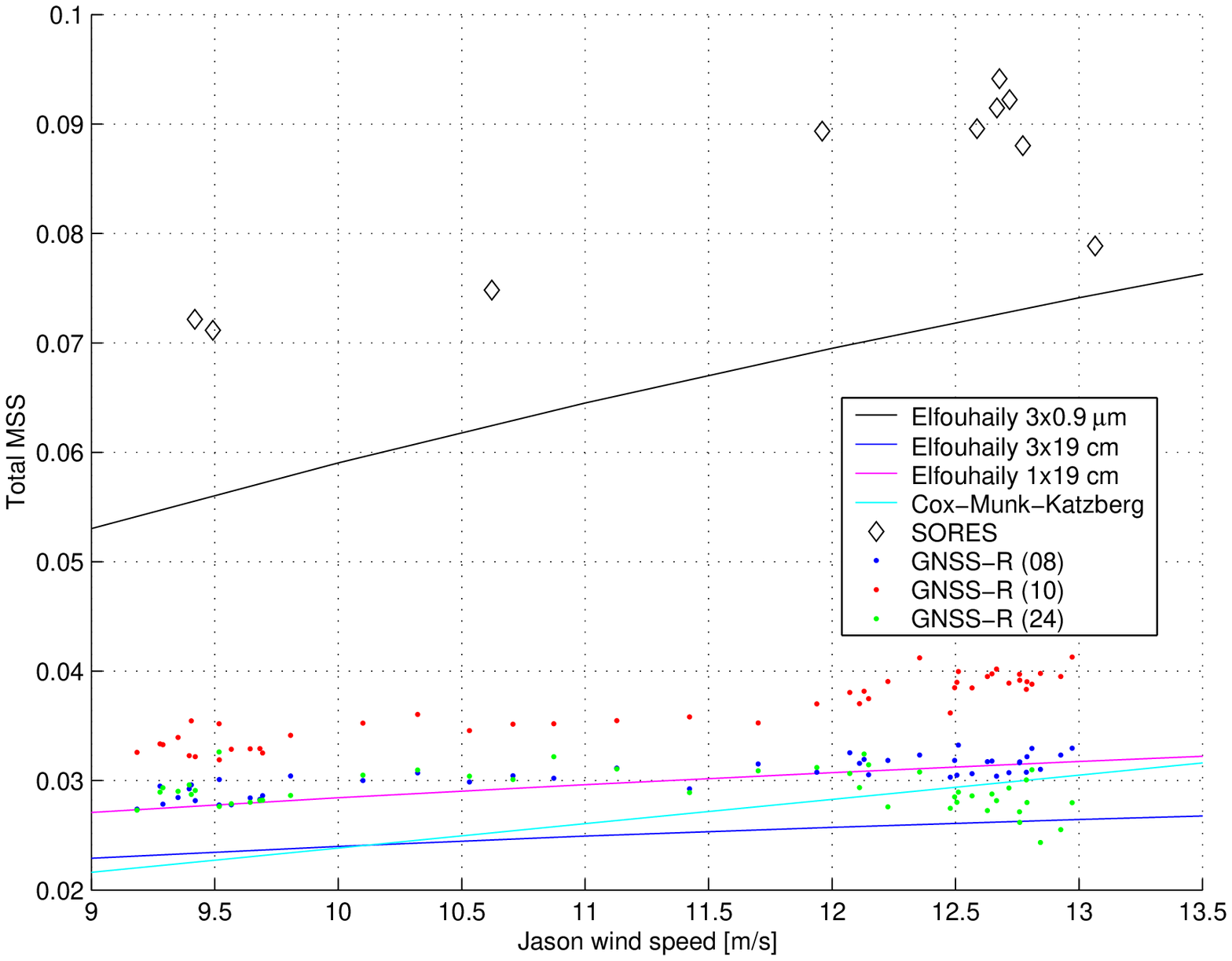}
\caption{
Total MSS versus Jason-1's wind speed.
}
\label{INTERPR1}
\end{figure*}

On Figure~\ref{INTERPR1}, we have plotted  the estimated total MSS versus Jason-1's wind speed
\footnote{The Jason-1 wind speed is derived from Ku-band $\sigma^o$ and significant wave height,
using the algorithm described in~\cite{gourrion2002a}.
Its stated accuracy is 2 m/s.}
together with two models:
\begin{itemize}
\item Elfouhaily's sea-height spectrum, integrated for different cut-off  wavelengths, and
\item an empirical model proposed by Katzberg for L-band, based on a modification of Cox and
Munk's relationship: MSS=$0.9.10^{-3}\sqrt{9.48U+6.07U^2}$, $U$ being wind speed (private communication with J.L. Garrison, Purdue University).
\end{itemize}
We see that both SORES and GNSS-R estimations follow Elfouhaily's model trend
(MSS obtained by integrating the spectrum with the usual cut-off of 3 times the wavelength) 
but give higher values of MSS (from to 20 to 40\% up-shifted).
Actually, we have found that MSS estimates of PRN08 and 24 are very well fitted by Elfouhaily's spectrum with a cut-off
of one wavelength only.
The 20\% discrepancy can be explained by a strong sea state with a SWH twice as high as the one observed during the Cox and Munk's experiment (almost 2~m compared to 1~m).
At any rate, these results indicate that the wind-MSS link is not straightforward and that the DMSS$_\lambda$ should be considered as a self-standing product for oceanographic users.

\section{Conclusion}

We have reported the first inversion of GNSS-R full Delay-Doppler Maps for the retrieval of the sea-surface directional mean square slope, DMSS$_{L}$. 
In addition, we have presented a repetition of the Cox and Munk experiment for DMSS$_{Opt}$ retrieval through inversion of Tilt Azimuth Map of  sun glitter optical photographs. 


Our results show that both optical and L-band total MSS are 20\% higher than  predicted using
Elfouhaily's model for the observed wind speed (9 to 13 m/s).
The SPA estimated by GNSS-R matches the swell direction with good accuracy for at least 2 out of 3 PRNs.
A new geophysical product
has been discussed: the slope PDF isotropy, which can be related to wind/wave misalignement as well the degree of sea development.

The analysis highlighted the important impact of sea-state unmodeled parameters (such as swell) in addition to wind stress on the measured DMSS$_{\lambda}$.
Since speculometry is sensitive to slope processes over a wide range of scales, the link between DMSS$_{\lambda}$ and wind is not straightforward: total MSS and SPA are definitely affected by swell. Quantitatively, the 20 \% bias observed in SORES results can be explained by the impact of swell on the elevation spectrum.
Consequently, DMSS$_{\lambda}$ can and should be studied as an independent parameter, of independent geophysical value.
We note however that the use of several wavelenghts could in principle be used to invert for all the parameters modulating the elevation spectrum, a line of future work.


Let us finally emphasize that the flight was not optimized for speculometry (1000~m altitude, 50~m/s speed) and that higher/faster flights are needed in the future in order to consolidate the concept of DDM inversion for DMSS$_\lambda$ estimation.

\section*{Acknowledgements}

This study was carried out under ESA contract 10120/01/NL/SF.
The dataset was collected in the frame of ESA contract TRP~ETP~137.A. We acknowledge all partners of the consortium (EADS-Astrium, Grupo de Mecanica del Vuelo, Institut d'Estudis Espacials de Catalunya and Institut Cartografic de Catalunya) for their contribution.

 {\em All Starlab authors have contributed significantly; the Starlab author list has been ordered randomly}.

\bibliographystyle{plain}
\addcontentsline{toc}{section*}{Bibliography}
\bibliography{/home/alkaid/intranet/library/feosbiblio.bib}
     
\end{multicols}
\end{document}